# Adsorption of CO and NO molecules on pristine, vacancy defected and doped graphene-like GaN monolayer: A first-principles study

Han-Fei Li[a,b,*], Si-Qi Li[a], Guo-Xiang Chen[a]

[a]*College of Sciences, Xi'an Shiyou University, Xi'an 710065, Shaanxi, PR China*

[b]*Key Laboratory of Weak-Light Nonlinear Photonics and School of Physics, Nankai University, Tianjin 300071, PR China*

**ABSTRACT**

In order to study the novel gas detection or sensing applications of gallium nitride monolayer (GaN-ML), we mainly focused on the structural, energetic, electronic and magnetic properties of toxic gas molecules (CO, NO) adsorbed on pristine, single vacancy (N-vacancy, Ga-vacancy) defected, and metals (Al, Fe, Pd and Pt) doped GaN-ML using density functional theory (DFT-D2 method) in this work. The calculations demonstrate that pristine GaN-ML is extremely insensitive to CO and NO together with the existence of a weak physisorption nature due to small adsorption energy, charge transfer, and long adsorption distance. It is found that both N-vacancy defected GaN-ML and Fe-doped GaN-ML can significantly increase the adsorption energy and charge transfer for CO. The CO adsorption induces the metallic characteristics of N-vacancy GaN-ML to be converted to the half-metallic characteristics together with 100% spin polarization, and it also drastically changes the magnetic moment, implying that N-vacancy GaN-ML exhibits excellent sensitivity to CO. However, Fe-doped GaN-ML is not conducive to CO detection. Moreover, N-vacancy defected and Pt-doped GaN-ML greatly improve the adsorption ability for NO compared to other substrates, and the presence of stronger orbital hybridization suggests that the interaction between them is chemisorption. Therefore, N-vacancy defected GaN-ML and Pt-doped GaN-ML can serve as potential materials in future NO sensing devices.

---

[*] Corresponding author.

 *E-mail address:* lihanfei2012@163.com.

# 1. Introduction

Stimulated by the pretty gigantic achievement of graphene [1], two-dimensional (2D) semiconductor materials, such as transition metal dichalcogenides and black phosphorene, have attracted considerable research interest from scientists due to their fascinating properties which can potentially be applied to nanoelectronic devices [2-11]. The experimental and theoretical studies have shown that the physical nature of atomic scale thickness endows the 2D semiconductors with more fruitful electronic, thermal, magnetic, optical and mechanical properties than those of corresponding bulk counterparts [12-18]. A novel 2D layered nanomaterial, namely graphene-like GaN monolayer (GaN-ML), exhibiting intriguing optoelectronic performance, high thermal and chemical stability, high specific surface area, and high surface activity, has been investigated intensively over the past decade [19-23]. Freeman et al. have predicted that graphene-like GaN-ML can transform into a 2D planar graphene-like structure when it is in the form of an ultrathin film [24]. Moreover, Şahin et al. have demonstrated that GaN-ML can form 2D stable nanostructures by phonon frequency spectrum calculations [25]. Recently, our previous paper has also reported that GaN-ML is dynamically stable via phonon calculations, and the calculated Ga–N bond length and cohesive energy are 1.87 Å and -7.85 eV, respectively [26]. Chen et al. reported that the 2D GaN-ML is a planar semiconductor with an indirect band gap of 1.95 eV using first-principles calculations a few years ago [27]. Al Balushi et al., Very recently, have synthesized 2D GaN-ML using a migration-enhanced encapsulated growth technique [28]. All of these studies indicate that GaN-ML is a vital 2D semiconductor material, which can be applied to nanoelectronics and optoelectronics. Therefore, it is worthwhile and valuable to theoretically investigate the stability and physical properties of 2D GaN-ML nanostructures for the future development of semiconductor nanomaterials.

It is well known that 2D nanomaterials have large surface-to-volume ratios and considerable changes of electric conductivity before and after gas adsorption, so they may be potential candidates for effective gas detection and sensing applications. The 2D GaN-ML overcomes the insurmountable obstacle of the zero band gap nature of graphene and the inherent defect of chemical inertness between the carbon atoms in the pristine graphene plane [29-30], which makes it have a stronger adsorption effect on most gases. This means that 2D GaN-ML is more conducive to the detection of gas molecules in contrast with the pristine graphene. However, it must be mentioned that the pristine GaN-ML is a non-magnetic material, which greatly hinders its application in spintronics and magnetic data storage. Hence, we hope to effectively modulate the spin characteristic of 2D GaN-ML by introducing heteroatom doping and defect structures, and it is expected to enhance the chemical reactivity of GaN-ML surface to improve the sensitivity for gas adsorption. For example, Tang et al. demonstrated that the magnetic properties of GaN-ML can be effectively tuned by the adsorption of non-metal atoms [31]. Recently, we have already predicated that Cr, Mn, Fe, Co and Ni doping can endow the GaN-ML with magnetization [32]. Li et al. predicted that Ga defective GaN-ML is magnetic and half-metallic, which may be an outstanding candidate for spintronics under certain conditions [33]. These studies show that the doping or vacancy defect may be a good way to modify the electronic and magnetic properties of 2D materials for exploring gas detection and sensing capabilities.

In this work, we present detailed studies of the adsorption properties of toxic gas molecules (CO, NO) on pristine, vacancy defected and doped GaN-ML using first-principles calculations. The adsorption energies, charges transfer, the band structures, density of states, charge density difference and magnetic properties of adsorbed systems are elaborated to confirm the high sensitivity and strong adsorption ability of the substrate. This work is intended to provide a theoretical guidance for experimenters to develop better 2D GaN-based materials for the preparation of high performance gas sensing devices.

## 2. Calculation methods and models

Our calculations are performed using the Vienna *ab initio* simulation package (VASP) based on density functional theory (DFT) [34-36]. The electron-ionic core interaction is represented by the projected augmented wave (PAW) potentials [37, 38], which are more accurate than the ultrasoft pseudopotentials. To treat exchange correlation interactions between electrons, we choose the Perdew–Burke–Ernzerhof (PBE) [39] formulation of the generalized gradient approximation (GGA), which yields the correct ground-state structure of the adsorbed systems. It is well known that GGA and local densities approximation (LDA) with DFT calculations are unable to describe correctly van der Waals (vdW) interactions. Specifically, GGA underestimates the relatively weak binding energies, whereas LDA overestimates them [40, 41]. Hence, the vdW interactions are described via the semi-empirical correction scheme of Grimme's DFT-D2 method [42] in our work. In order to model gas molecules adsorbed on pristine, vacancy defected and doped GaN-ML, we choose a $4 \times 4 \times 1$ supercell in the simulation cell. The distance between adjacent layers is kept as 18 Å along the z-axial direction, which is sufficient to eliminate the interaction between periodic images. A plane-wave basis set with kinetic energy cutoff of 500 eV is utilized, and the threshold for structure relaxation and self-consistency are set as 0.02 eV/Å and $10^{-5}$ eV, respectively. Integration over the Brillouin zone is performed by using the Gamma-centered Monkhorst-Pack scheme [43] with (11 × 11 × 1) k-points, while the total density of states (DOS) is obtained with a ten-time denser mesh and a Gaussian smearing broadening of 0.1 eV.

The energetically more favorable sites are determined by the calculated adsorption energy ($E_{ad}$) of the gas molecules adsorbed GaN-ML, which is defined as $E_{ad} = E_{total} - E_{sheet} - E_{gas}$, Where $E_{total}$, $E_{sheet}$ and $E_{gas}$ are the total energy of pristine, vacancy defected or doped GaN-ML adsorbed system, pristine GaN-ML, vacancy defected GaN-ML or doped GaN-ML, and an isolated gas molecule, respectively. A negative $E_{ad}$ corresponds to a stable adsorption structure. The calculated adsorption energy and the amount of charge transfer obtained by Bader charge analysis [44] are listed in Table 1 and Table 2, together with the corresponding structural parameters, including adsorption distance ($D$) and molecular bond length ($L_{C/N-O}$). $D$ is defined as the distance of nearest atoms between the substrate and gas molecules. In addition, $Q$ is the charge transfer between the substrate and gas molecules, and a negative value means charge transfer from the substrate to the gas molecule.

## 3. Results and Discussions

*3.1. Stability and geometric structures*

In this paper, in order to obtain the most energetically stable adsorption configurations between the substrate and gas molecules, the gas molecule is initially placed at different positions above GaN-ML with different orientations. As shown in Fig. 1, four possible initial adsorption sites of gas molecules are considered, namely, top of Ga atoms ($T_{Ga}$), top of N atoms ($T_N$), top of the Ga-N bond center (B), and top of the hexagon center (H). For gas molecules (CO, NO), three initial molecular orientations are tested at each adsorption site, i.e., the molecular is parallel to the substrate plane, while the other two are perpendicular to it. Concretely speaking, the CO molecule is perpendicular to the substrate, which means that the C atom points to the substrate and away from it, as is the case with NO. The optimized adsorption values of the most stable adsorption configurations of the CO and NO adsorbed on pristine GaN-ML are summarized in Table 1. We can see that the values of adsorption energy for CO adsorbed on pristine GaN-ML at the four adsorption sites are almost the same, indicating that CO is not sensitive to these adsorption sites of pristine GaN-ML. Compared with the other three sites, since there are relatively larger adsorption energies (-0.16 eV) and charge transfer (0.04 $e$) at H site, we take the H site as an example of the most stable configuration for CO adsorbed on pristine GaN-ML (see Fig. 2(a)). Similar to the case of CO, it can be clearly seen that the values of adsorption energy and charge transfer at the four adsorption sites are almost the same for the NO adsorbed system, suggesting that NO adsorption is independent on these adsorption sites. Due to there are the lowest adsorption energy of -0.37 eV, the shortest adsorption distance of 2.46 Å and the most charge transfer of 0.17 $e$ from pristine GaN-ML to $H_2S$ at B site, we take the B site as an example of the most stable adsorption site (see Fig. 2(b)). As we have seen, the C atom of the CO points towards the substrate surface and the O atom points away from it, which is one stable configuration, but the substrate is not induced to undergo any structural deformation. Moreover, the original C–O bond length of 1.143 Å is almost constant after CO adsorption. For the case of NO, the original N–O bond length of 1.169 Å is elongated to 1.18 Å after NO adsorption, indicating that the N–O bond is basically unchanged. In addition, we find that the NO molecule with the lower N atom sitting above the B site does not cause obvious structural distortions of the substrate. These mean that CO and NO molecules all are physisorbed on pristine GaN-ML with small adsorption energy, charge transfer, and the larger adsorption distance, indicating that pristine GaN-ML is insensitive to them.

To significantly enhance the sensitivity of GaN-ML to gas molecules, we investigate the effects of N-vacancy defect, Ga-vacancy defect, and metal (Al, Fe, Pd, Pt) doping on the adsorption ability of GaN-ML for gas molecules. The gas molecules are initially placed on the top of the vacancies and doping sites with different molecular orientations similar to the case of the $T_{Ga}$ site described above. The most stable adsorption configurations of CO and NO molecules on vacancy defected and metal doped GaN-ML are plotted in Fig. 3. It is obvious that the metal atoms induce the structural deformations of the six-membered ring near the doping site, causing metal atoms to protrude or dent in varying degrees (see Fig. 3(c)-(f) and (i)-(l)). For the case of the vacancy defect, adsorption of CO and NO results in the Ga atom near the N-vacancy defect protruding out of the GaN-ML surface, and the N atom near the Ga-vacancy defect is lifted high (see Fig. 3(a)-(b) and (g)-(h)). These phenomena mean that there are the strong interactions between the gas molecules and the substrate. From Table 2, we can find that the N-vacancy defect significantly enhances the adsorption ability of GaN-ML for CO compared to other substrates, together with the lowest adsorption energy (-0.95 eV), the shortest adsorption distance (2.03 Å) and the most charge transfer (1.13 $e$). The C–O bond length is elongated from 1.143 Å to 1.32 Å in the N-vacancy defected adsorption system. Besides, for doping, Fe doped GaN-ML significantly enhance adsorption energy (-0.60 eV) and charge transfer (0.13 $e$) of GaN-ML for CO compared to other metal dopants. It is interesting to note that CO adsorbed on Pd and Pt doped GaN-ML have larger adsorption energy compared with the pristine GaN-ML, but less charge transfer implies that they are not conducive to CO detection. For the adsorption of NO, NO adsorbed on N-vacancy defected GaN-ML has the lowest adsorption energy (-1.61 eV) and the most charge transfer (0.57 $e$) compared with the Ga-vacancy defected GaN-ML, indicating that the N-vacancy defect is more conducive to NO detection. The N–O bond length increases to 1.23 Å from the original 1.169 Å by introducing the N-vacancy defect. Obviously, Pt doped GaN-ML is the best way to enhance the adsorption ability of GaN-ML for NO in all doping systems, which is attributed to the lowest adsorption energy (-1.94 eV), the most charge transfer (0.32 $e$) and significantly decreased adsorption distance (1.85 Å). NO adsorbed on Pt doped GaN-ML brings about a serious structural deformation of the substrate, that is, the Pt atom is pulled up to a height of 1.03 Å as shown in Fig. 3(l), thereby indicating that the interaction between NO and Pt doped GaN-ML is extremely strong and belongs to chemisorption. Undoubtedly, our calculated results confirm that the vacancy defect and doping can enhance the adsorption ability of GaN-ML for CO and NO.

*3.2. Band structure and densities of states*

For gaining a better insight into the effects of gas molecules adsorption on the electronic properties of pristine GaN-ML, we analyze the total and partial densities of states (TDOS and PDOS) and band structures of the adsorbed systems. The band structures of gas molecules adsorbed on pristine GaN-ML based on the most stable configuration are shown in Fig. 4. The black lines and red lines represent the spin-up and spin-down components, respectively. The Fermi level is set to zero energy and indicated by the black horizontal dashed line. We can clearly observe from Fig. 4(a) and (b) that the spin-up and spin-down band channels are symmetrical, indicating that the existence of spin degeneracy, which means that the pristine GaN-ML with and without CO adsorption are proved to be nonmagnetic nature. However, the spin-up and spin-down band channels are asymmetric in Fig. 4(c) due to the presence of two impurity bands near the Fermi level, which indicates that spin degeneracy is broken, revealing the presence of a net magnetization after NO adsorption. It is reported that the band gap is one of the vital parameters to determine the electrical conductivity of materials, and there is a classic relationship between the electric conductivity (σ) and the band gap ($E_g$) is as follows [45]:

$$\sigma \propto \exp(\frac{-E_g}{2KT}) \qquad (1)$$

where $K$ and $T$ refer to the Boltzmann's constant and the Kelvin temperature, respectively. Apparently, a small change in $E_g$ can lead to a considerable change in σ. According to our calculations, pristine GaN-ML has indirect band gap of 2.04 eV, which is almost the same as the band gap of the CO adsorbed system, indicating that there is no an evident change in electric conductivity. This further confirms that pristine GaN-ML is not sensitive to CO. The TDOS of pristine GaN-ML with and without gas molecules adsorption as well as PDOS projected to the gas adatom and its neighboring substrate atoms are shown in Fig. 5. For each graph, the positive and negative DOS represent spin-up and spin-down states, respectively. The Fermi level is set to zero energy and indicated by the black vertical dashed lines. It can be clearly seen from Fig. 5(a) and (b) that the pristine GaN-ML with and without CO adsorption are a nonmagnetic (NM) semiconductor, but two impurity states appear near the Fermi level after NO adsorption, indicating that NO adsorption induces spin polarization. Moreover, the hybridization between the 2p orbitals of CO and NO and the 2p and 4p orbitals of the substrate atoms is almost invisible from PDOS, implying that pristine GaN-ML is insensitive to them. This further indicates that CO and NO have low reactivity to pristine GaN-ML in vicinity of the adsorption sites.

It is necessary and indispensable to further explore the effects of gas molecules adsorption on the electronic properties and magnetic behaviors of the various defected GaN-ML including the vacancy defect and substitutional doping, thereby we next investigate the band structures, the TDOS of the various defected GaN-ML with and without gas molecules adsorption as well as PDOS projected on the gas adatom and its neighboring substrate atoms. As we can see, the band structures of CO and NO adsorbed on the various defected GaN-ML are also depicted in Fig. 6. For CO adsorption, only CO adsorbed on Al doped GaN-ML exhibits NM properties due to spin-up and spin-down band channels are symmetrical (see Fig. 6(c)). It is noted from Fig. 6(a) that one band channel is tangent to the Fermi level at the Γ point in spin-down bands, while spin-up bands present characteristics of the semiconductor, revealing the occurrence of the half-metallic characteristics in the N-vacancy adsorbed system. The remaining cases are all representative of the semiconducting properties, which is due to the Fermi level being in the gap of between the valence band maximum (VBM) and the conduction band minimum (CBM) for both spin-up and spin-down band channels (see Fig. 6(b) and (d)-(f)). In the cases of NO adsorbed systems, Pd and Pt doped GaN-ML all present the NM semiconducting properties after NO adsorption, while other cases indicate the existence of magnetic semiconductor characteristics (see Fig. 6(g)-(l)). It is worth mentioning that the total magnetic moments of the adsorption systems are mainly contributed by the metal dopants (see Table 2). However, the total magnetic moment of the 1.00 $\mu_B$ in the NO adsorbed Al doped GaN-ML is completely contributed by NO.

TDOS and PDOS of CO and NO adsorbed on the various defected GaN-ML are displayed in Fig. 7. From the DOS of the CO adsorbed N-vacancy GaN-ML plotted in Fig. 7(a), we can see that the contribution of CO electronic levels to the TDOS of the adsorbed systems appears near the Fermi level from -0.36 to 0.34 eV and below the Fermi level from -0.90 to -0.18 eV compared with N-vacancy GaN-ML. After CO adsorption, there is DOS peaks for the new occupied states appearing between -0.90 to -0.18 eV, which are mainly contributed by 2p states of the C atom and O atom and 4p states of the nearest-neighbor Ga atoms of the adsorption site according to the PDOS analysis. Besides, the orbitals of the C, O and Ga atoms possess a large number of similar states, indicating that the emergence of a strong orbital hybridization. Considering the larger adsorption energy and charge transfer, we further confirm that CO adsorbed on N-vacancy GaN-ML belongs to chemisorption, meaning that the N-vacancy GaN-ML substrate is beneficial for CO detection. It is worth mentioning that CO adsorption causes the metallic characteristics of N-vacancy GaN-ML to be converted to the half-metallic characteristics, indicating a distinct change in conductivity. And CO adsorption significantly increases the magnetic moment of the substrate from original 0.00 $\mu_B$ to 1.00 $\mu_B$. These demonstrate that N-vacancy GaN-ML has excellent sensitivity to CO and increases the reactivity of the GaN-ML surface toward the adsorption of CO. For CO adsorbed on Fe doped GaN-ML as shown in Fig. 7(d), it can be clearly seen that the 2p states of the C and O atom in CO and 3d states of the Fe atom are less overlapped near 1.87 eV and 2.5 eV, which means that the occurrence of a weak orbital hybridization. Moreover, the total magnetic moment of Fe doped GaN-ML is 5.00 $\mu_B$/unit cell, which is the same as the total magnetic moment after CO adsorption. At the same time, the adsorption of CO allows the semiconducting properties of Fe doped GaN-ML to be retained. Although there is larger adsorption energy and charge transfer between Fe doped GaN-ML and CO, the constant electronic properties and magnetic moment changes after CO adsorption imply that Fe doped GaN-ML is not conducive to CO detection. There is no obvious orbital hybridization in other cases of CO adsorption as shown in PDOS plots (see Fig. 7(b)-(c) and (e)-(f)). In the case of NO adsorbed on N-vacancy GaN-ML, one can see that the orbitals of the N, O and Ga atoms possess a great deal of similar states in the region between -0.85 to 1.8 eV shown in Fig. 7(g), indicating that the emergence of a strong orbital hybridization and it belongs to chemisorption. After NO adsorption, N-vacancy GaN-ML is converted into the semiconducting properties from the metallic characteristics, so the conductivity is significantly weakened. And the adsorption of NO induces the magnetic moment of the adsorbed system to remarkably increase to 2.00 $\mu_B$. Hence, the N-vacancy GaN-ML facilitates the detection of NO. Meanwhile, N-vacancy GaN-ML also shows high reactivity toward NO. As shown in Fig. 7(l), it is found that the 2p states of the N and O atom in NO and 5d states of the Pt atom are more overlapped near -0.6 eV and 1.5 eV, which means that the occurrence of the strong orbital hybridization, indicating the interaction between NO and Pt doped GaN-ML is a chemisorption. In addition, after NO adsorption, the electronic properties of the Pt doped GaN-ML change from the original half-metallic properties to the semiconducting properties, and the magnetic moment changes drastically from original 1.00 $\mu_B$ to 0.00 $\mu_B$. Therefore, Pt doped GaN-ML is the most effective method to enhance the sensitivity of GaN-ML for NO compared with other metal doping (see Fig. 7(i)-(l)). This also proves Pt doped GaN-ML increases the reactivity of the GaN-ML surface to NO.

*3.3. Charge density difference*

Through the calculation and analysis of the charge density difference [46] (CDD), we can clearly see the electron transfer and charge density distribution, which can facilitate us to further understand the electronic properties of the interaction between gas molecules and the various defected GaN-ML. The CDD images of CO and NO adsorbed on the various defected GaN-ML are shown in Fig. 8, calculated by the formula $\Delta\rho = \rho_{total} - \rho_{sheet} - \rho_{gas}$, where $\rho_{total}$, $\rho_{sheet}$ and $\rho_{gas}$ are the charge density of gas molecules adsorbed on the various defected GaN-ML, the various defected GaN-ML and gas molecules, respectively. $\rho_{sheet}$ and $\rho_{gas}$ must be calculated in the same adsorbed structure. As shown in Fig. 8(a), the chemisorption can be confirmed for CO adsorbed N-vacancy GaN-ML. The CDD image shows that the charge accumulation mainly occurs in the region between CO and the substrate while charges dissipate at the top of the C atom. And there is a charge accumulation appearing on both sides of the O atom in CO. In addition, the charge dissipation is found on the nearest-neighbor Ga atom at the adsorption site in the substrate. The Bader charge analysis suggests that CO acts as an electron acceptor with more charge transfer from the substrate to CO of 1.13 $e$. These confirm that N-vacancy GaN-ML significantly increases the reactivity to CO near the adsorption site. The CDD of CO adsorbed on Fe doped GaN-ML is shown in Fig. 8(d), in which it can be found that the charge accumulation mainly occurs in the region between the C atom and the Fe atom and almost envelops the CO molecule. The charge dissipation emerges above the Fe atom, while there is charge accumulation on the nearest-neighbor N atom at the doping site. This confirms that the CO adsorbed Fe doped GaN-ML gives rise to a chemisorption. The Bader charge analysis shows that the CO gains 0.13 $e$ from the Fe doped GaN-ML and acts as an electron acceptor. We can also visually see that CO adsorbed on Ga-vacancy defected GaN-ML, Pd doped GaN-ML and Pt doped GaN-ML are a feeble physisorption (see Fig. 8(b) and (e)-(f)). In the case of NO adsorbed N-vacancy GaN-ML shown in Fig.8 (g), the charge depletion mainly occurs in the region between the NO and the substrate while the charge accumulation surrounds the N and O atom. The NO obtains 0.57 $e$ from N-vacancy GaN-ML by the Bader charge analysis. For NO adsorbed on Pt doped GaN-ML as shown in Fig. 8(l), it can be found that an obvious charge piling appears in the region between the N atom and the nearest-neighbor Pt atom of the adsorption site. And the charge accumulation surrounds the O atom while the charges dissipate on both sides of the O atom. Therefore, the chemisorption of NO adsorbed on Pt-GaN is verified. According to Bader charge analysis, NO has obtained charges from the Pt doped GaN-ML by 0.32 $e$ and thus NO acts as an electron acceptor. Moreover, it can be seen that the Pt doped GaN-ML significantly increases the reactivity of GaN-ML for NO near the doping sites compared with other metal doping (see Fig. 8(i)-(l)).

## 4. Conclusions

In summary, the structural, energetic, electronic and magnetic properties of the pristine GaN-ML,

vacancy defected and doped GaN-ML with the adsorbed toxic gas molecules (CO, NO) have been investigated using the first-principles method. The DFT-D2 calculations unravel that pristine GaN-ML is extremely insensitive to CO and NO together with the existence of a weak physisorption nature due to small adsorption energy, charge transfer, and long adsorption distance. These demonstrate that CO and NO have low reactivity to pristine GaN-ML. Both N-vacancy defected GaN-ML and Fe doped GaN-ML significantly increase the adsorption energy and charge transfer for CO adsorption. The CO adsorption causes the metallic characteristics of N-vacancy GaN-ML to be converted to the half-metallic characteristics together with 100% spin polarization, and CO adsorption drastically increases the magnetic moment, indicating that N-vacancy GaN-ML exhibits excellent sensitivity to CO. However, Fe doped GaN-ML is not conducive to CO detection owing to the constant electronic properties and magnetic moment changes after CO adsorption. Moreover, N-vacancy GaN-ML and Pt doped GaN-ML greatly enhance the adsorption ability of GaN-ML for NO as well as the reactivity of the GaN-ML surface to NO compared to other substrates. The presence of the strong orbital hybridization suggests that the interaction between them is chemisorption. After NO adsorption, the original electronic properties of N-vacancy GaN-ML and Pt doped GaN-ML are converted into the semiconducting properties with a pronounced change in magnetic moment. Therefore, N-vacancy defected GaN-ML and Pt doped GaN-ML can serve as potential materials in future NO sensing devices.

## Acknowledgements

This work is supported by the National Natural Science Foundation of China (Grant no. 11304246), the Shaanxi Province Science and Technology Foundation (Grant no. 2014KJXX-70), and the Postgraduate Innovation and Practical Ability Training Program of Xi'an Shiyou University (Grant no. YCS17111020).

**Table Captions**

Table 1. CO and NO adsorbed on four different adsorption sites of pristine GaN-ML: adsorption energy ($E_{ad}$), adsorption distance ($D$) of nearest-neighbor atoms between gas molecule and substrate, bond length ($L_{C/N-O}$) for the different adsorption sites of CO or NO adsorbed on pristine GaN-ML, and charge transfer ($Q$) between gas molecule and substrate.

Table 2. CO and NO adsorbed on various defected GaN-ML: adsorption energy ($E_{ad}$), adsorption distance ($D$) of nearest-neighbor atoms between gas molecule and various defected substrate, bond length ($L_{C/N-O}$) of CO or NO adsorbed on various defected GaN-ML, total magnetic moment ($\mu_{tot}$) of the whole adsorption systems, local magnetic moment ($\mu_M$) on the metal atoms, and charge transfer ($Q$) between gas molecule and various defected substrate.

**Figure Captions**

Fig. 1. The optimized geometry of pristine GaN-ML and four possible adsorption sites including top of Ga atoms (T$_{Ga}$), top of N atoms (T$_N$), top of the Ga-N bond center (B), and top of the hexagon center (H).

Fig. 2. Top and side views of the most stable adsorption configurations of (a) CO, (b) NO adsorbed on pristine GaN-ML surface. The C and O atoms are shown in gray and red balls, respectively.

Fig. 3. Top and side views of the most stable adsorption configurations of CO ((a)-(f)) and NO ((g)-(l)) adsorbed on the N-vacancy, Ga-vacancy, Al, Fe, Pd and Pt doped GaN-ML. The Al, Fe, Pd and Pt atoms are shown in pink, purple, cyan and green balls, respectively.

Fig. 4. The band structures of (a) pristine GaN-ML and (b) CO, (c) NO adsorbed on pristine GaN-ML based on the most stable configuration. The black lines and red lines represent the spin-up and spin-down components, respectively. The Fermi level is set to zero energy and indicated by the black horizontal dashed line.

Fig. 5. The total DOS and PDOS of (a) CO, (b) NO adsorbed on pristine GaN-ML based on the most stable configuration. The positive (negative) values of DOS denote the spin-up (spin-down) states. The Fermi level is set to zero energy and indicated by the black vertical dashed line.

Fig. 6. The band structures of CO ((a)-(f)) and NO ((g)-(l)) adsorbed on the N-vacancy, Ga-vacancy, Al, Fe, Pd and Pt doped GaN-ML based on the most stable configuration. The Fermi level is set to zero energy and indicated by the black horizontal dashed line.

Fig. 7. The total DOS and PDOS of CO ((a)-(f)) and NO ((g)-(l)) adsorbed on the N-vacancy, Ga-vacancy, Al, Fe, Pd and Pt doped GaN-ML based on the most stable configuration. The Fermi level is set to zero energy and indicated by the black vertical dashed lines.

Fig. 8. Top and side views of the charge density difference for CO ((a)-(f)) and NO ((g)-(l)) adsorbed on the N-vacancy, Ga-vacancy, Al, Fe, Pd and Pt doped GaN-ML based on the most stable configuration. The prunosus and yellow regions correspond to charge accumulation and depletion, respectively. The isosurface is taken as 0.003 e/Å$^3$ for (a) and (g)-(l); 0.001 e/Å$^3$ for (b)-(d); 0.0002 e/Å$^3$ for (e) and (f).

**Table 1**

CO and NO adsorbed on four different adsorption sites of pristine GaN-ML: adsorption energy ($E_{ad}$), adsorption distance ($D$) of nearest-neighbor atoms between gas molecule and substrate, bond length ($L_{C/N-O}$) for the different adsorption sites of CO or NO adsorbed on pristine GaN-ML, and charge transfer ($Q$) between gas molecule and substrate.

| Molecule | Site | $E_{ad}$ (eV) | $D$ (Å) | $L_{C/N-O}$ (Å) | $Q$ (e) |
|---|---|---|---|---|---|
| CO | $T_{Ga}$ | -0.13 | 3.04 | 1.14 | -0.02 |
|    | $T_N$    | -0.15 | 3.04 | 1.15 | -0.04 |
|    | B        | -0.15 | 3.01 | 1.14 | -0.02 |
|    | H        | -0.16 | 3.11 | 1.15 | -0.04 |
| NO | $T_{Ga}$ | -0.34 | 2.59 | 1.18 | -0.16 |
|    | $T_N$    | -0.35 | 2.60 | 1.18 | -0.16 |
|    | B        | -0.37 | 2.46 | 1.18 | -0.17 |
|    | H        | -0.36 | 2.47 | 1.18 | -0.16 |

**Table 2**

CO and NO adsorbed on various defected GaN-ML: adsorption energy ($E_{ad}$), adsorption distance ($D$) of nearest-neighbor atoms between gas molecule and various defected substrate, bond length ($L_{C/N-O}$) of CO or NO adsorbed on various defected GaN-ML, total magnetic moment ($\mu_{tot}$) of the whole adsorption systems, local magnetic moment ($\mu_M$) on the metal atoms, and charge transfer ($Q$) between gas molecule and various defected substrate.

| Systems | $E_{ad}$ (eV) | $D$ (Å) | $L_{C/N-O}$ (Å) | $\mu_{tot}$ ($\mu_B$) | $\mu_M$ ($\mu_B$) | $Q$ (e) |
|---|---|---|---|---|---|---|
| $Ga_{16}N_{15}$ + CO | -0.95 | 2.03 | 1.32 | 1.00 | – | -1.13 |
| $Ga_{15}N_{16}$ + CO | -0.40 | 2.85 | 1.15 | 3.00 | – | -0.05 |
| $Ga_{15}N_{16}Al$ + CO | -0.37 | 2.17 | 1.15 | 0.00 | 0.00 | -0.14 |
| $Ga_{15}N_{16}Fe$ + CO | -0.60 | 2.04 | 1.15 | 5.00 | 3.47 | -0.13 |
| $Ga_{15}N_{16}Pd$ + CO | -0.64 | 3.27 | 1.15 | 1.00 | 0.41 | -0.06 |
| $Ga_{15}N_{16}Pt$ + CO | -0.61 | 3.31 | 1.15 | 1.00 | 0.46 | -0.05 |
| $Ga_{16}N_{15}$ + NO | -1.61 | 2.03 | 1.23 | 2.00 | – | -0.57 |
| $Ga_{15}N_{16}$ + NO | -1.03 | 2.33 | 1.14 | 2.00 | – | 0.27 |
| $Ga_{15}N_{16}Al$ + NO | -0.50 | 2.12 | 1.18 | 1.00 | 0.00 | -0.26 |
| $Ga_{15}N_{16}Fe$ + NO | -1.41 | 1.63 | 1.19 | 2.00 | 1.83 | -0.36 |
| $Ga_{15}N_{16}Pd$ + NO | -1.67 | 1.91 | 1.19 | 0.00 | 0.00 | -0.22 |
| $Ga_{15}N_{16}Pt$ + NO | -1.94 | 1.85 | 1.20 | 0.00 | 0.00 | -0.32 |

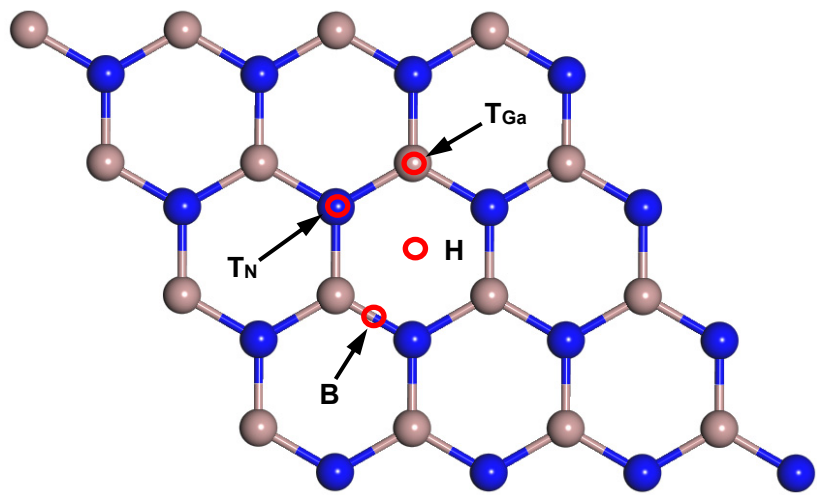

**Fig. 1.** The optimized geometry of pristine GaN-ML and four possible adsorption sites including top of Ga atoms ($T_{Ga}$), top of N atoms ($T_N$), top of the Ga-N bond center (B), and top of the hexagon center (H).

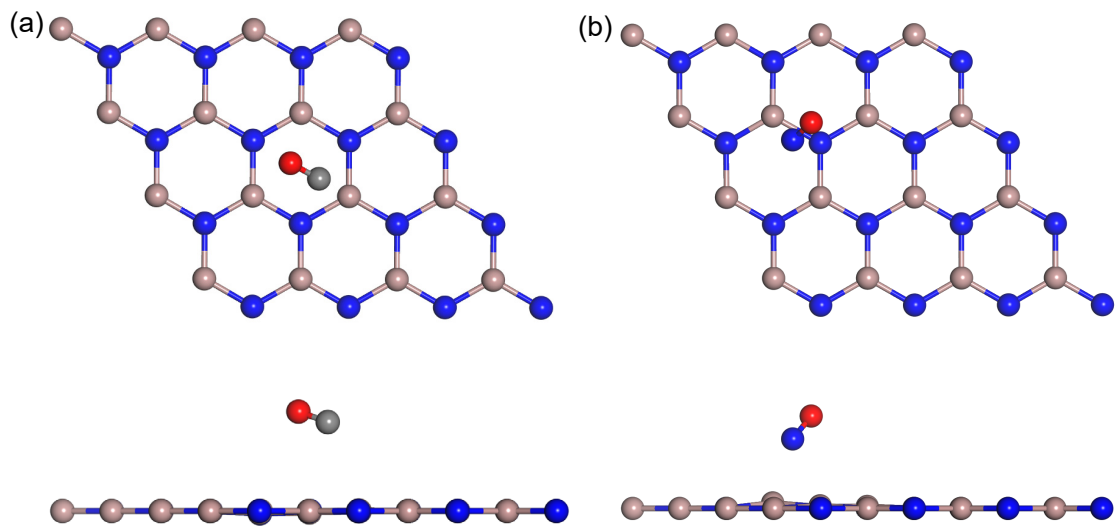

**Fig. 2.** Top and side views of the most stable adsorption configurations of (a) CO, (b) NO adsorbed on pristine GaN-ML surface. The C and O atoms are shown in gray and red balls, respectively.

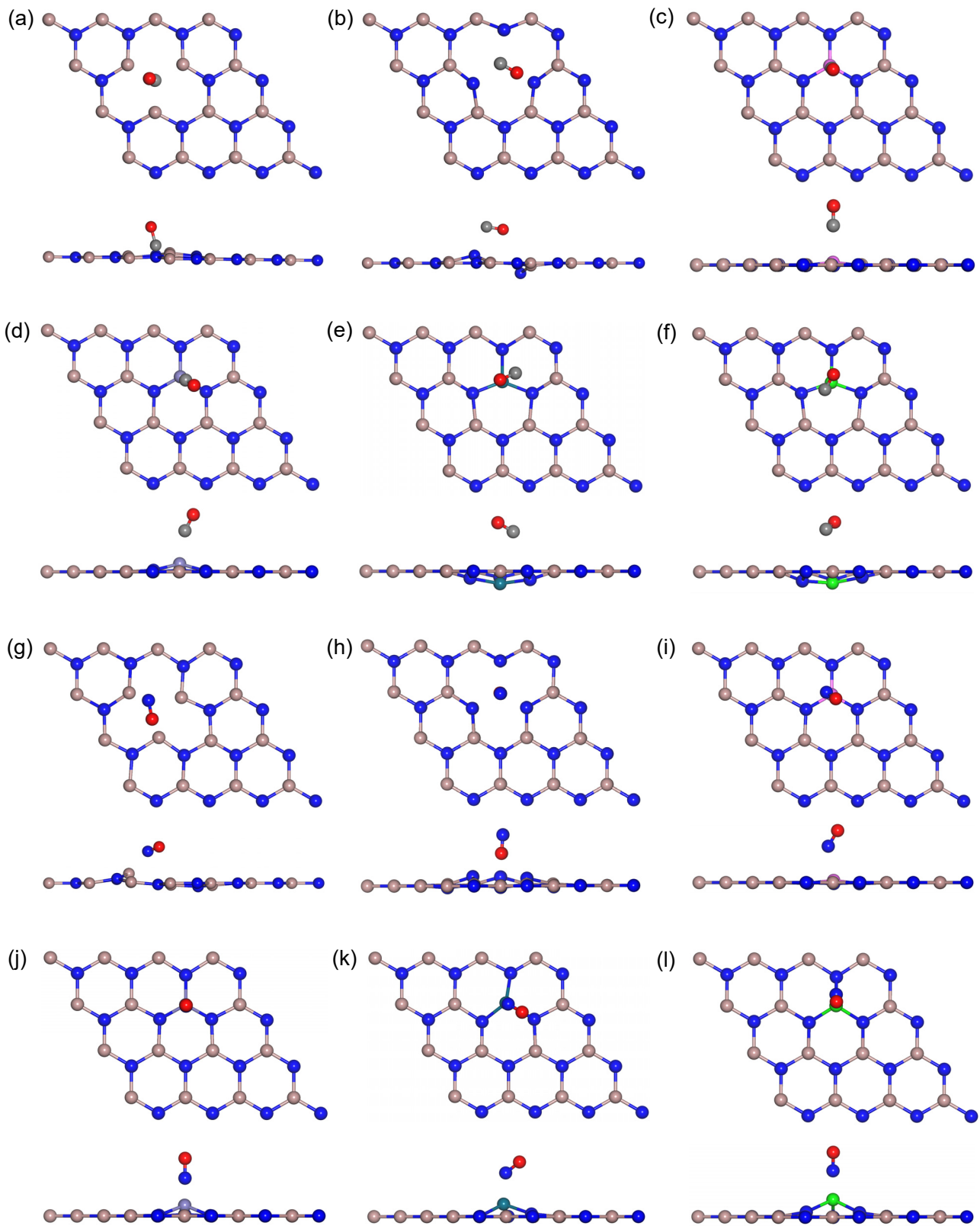

**Fig. 3.** Top and side views of the most stable adsorption configurations of CO ((a)-(f)) and NO ((g)-(l)) adsorbed on the N-vacancy, Ga-vacancy, Al, Fe, Pd and Pt doped GaN-ML. The Al, Fe, Pd and Pt atoms are shown in pink, purple, cyan and green balls, respectively.

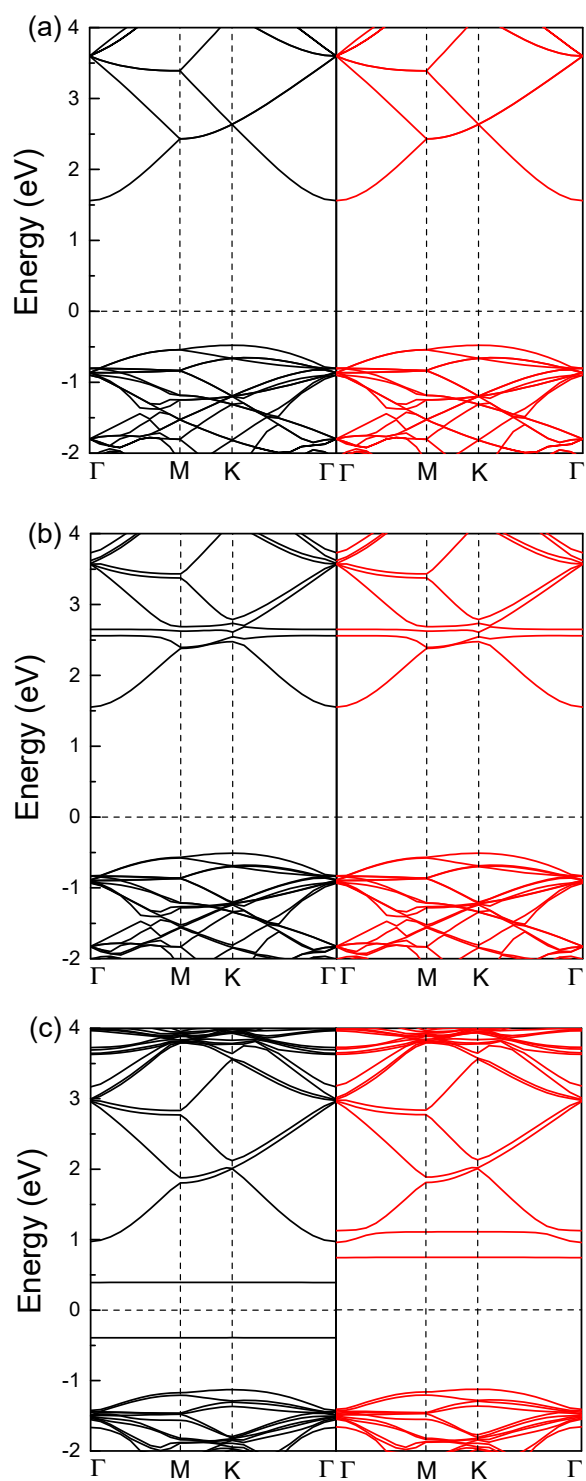

**Fig. 4.** The band structures of (a) pristine GaN-ML and (b) CO, (c) NO adsorbed on pristine GaN-ML based on the most stable configuration. The black lines and red lines represent the spin-up and spin-down components, respectively. The Fermi level is set to zero energy and indicated by the black horizontal dashed line.

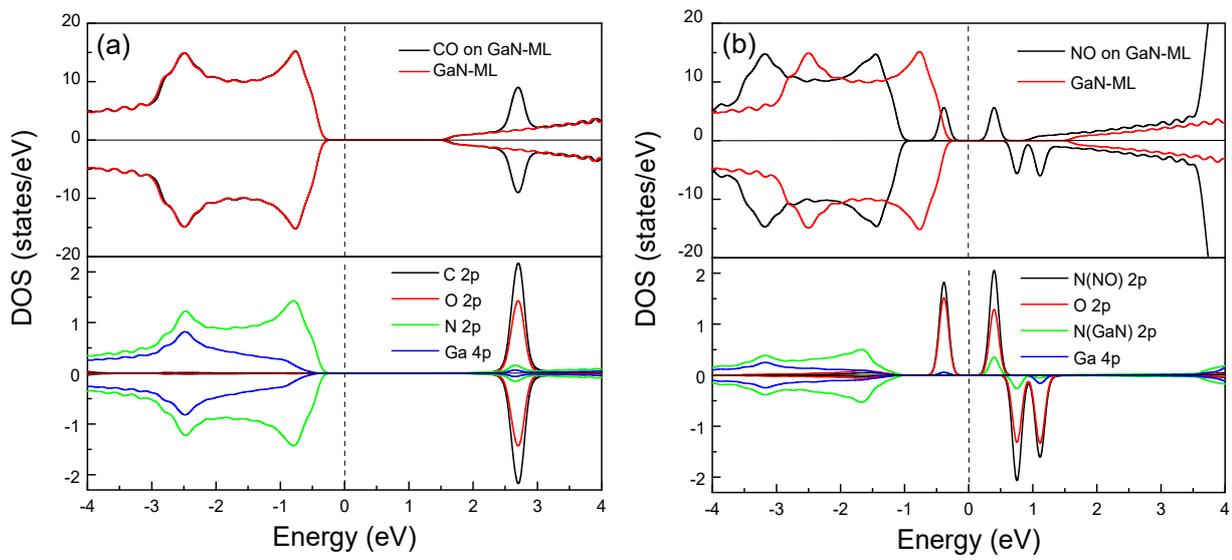

**Fig. 5.** The total DOS and PDOS of (a) CO, (b) NO adsorbed on pristine GaN-ML based on the most stable configuration. The positive (negative) values of DOS denote the spin-up (spin-down) states. The Fermi level is set to zero energy and indicated by the black vertical dashed line.

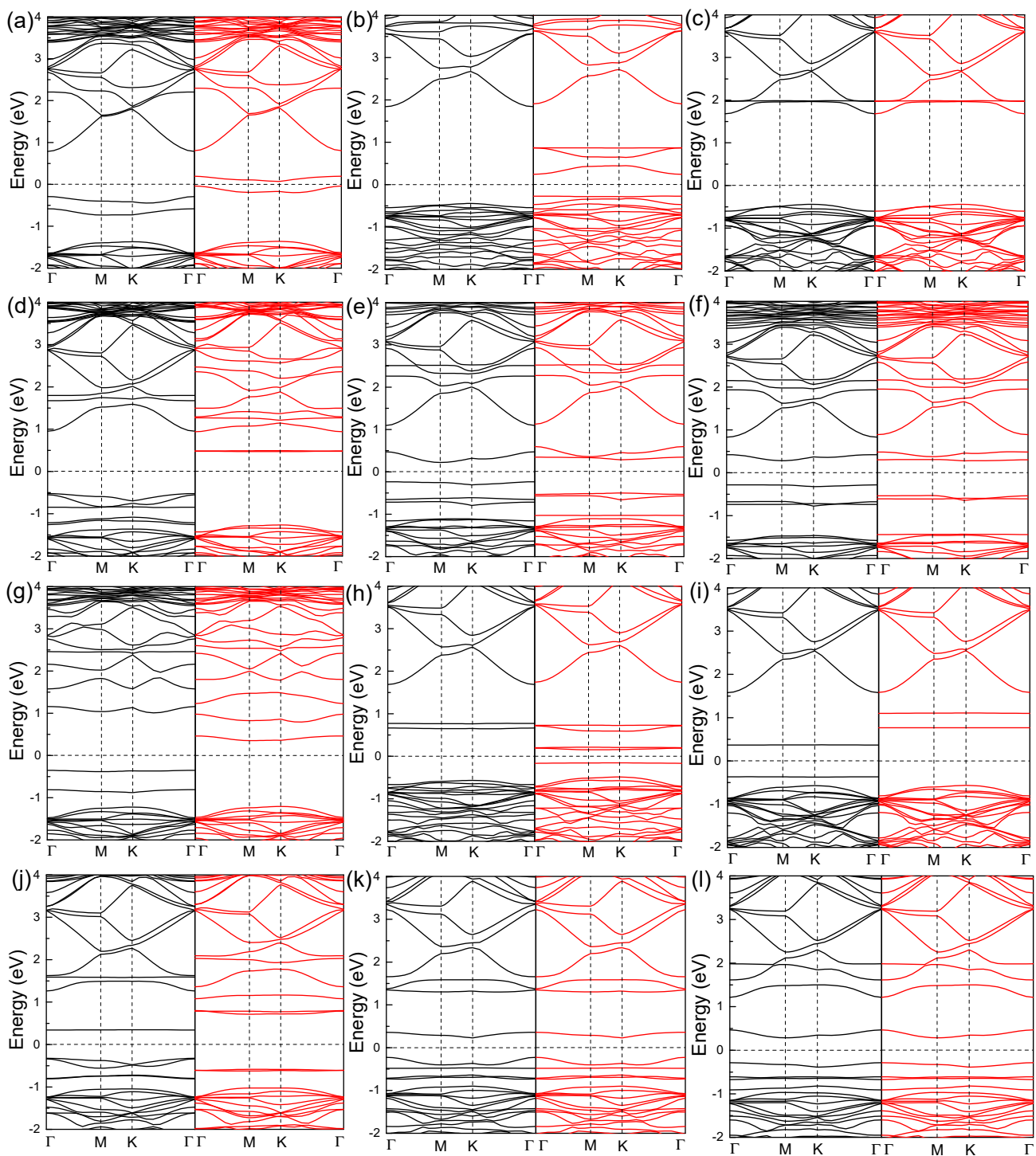

**Fig. 6.** The band structures of CO ((a)-(f)) and NO ((g)-(l)) adsorbed on the N-vacancy, Ga-vacancy, Al, Fe, Pd and Pt doped GaN-ML based on the most stable configuration. The Fermi level is set to zero energy and indicated by the black horizontal dashed line.

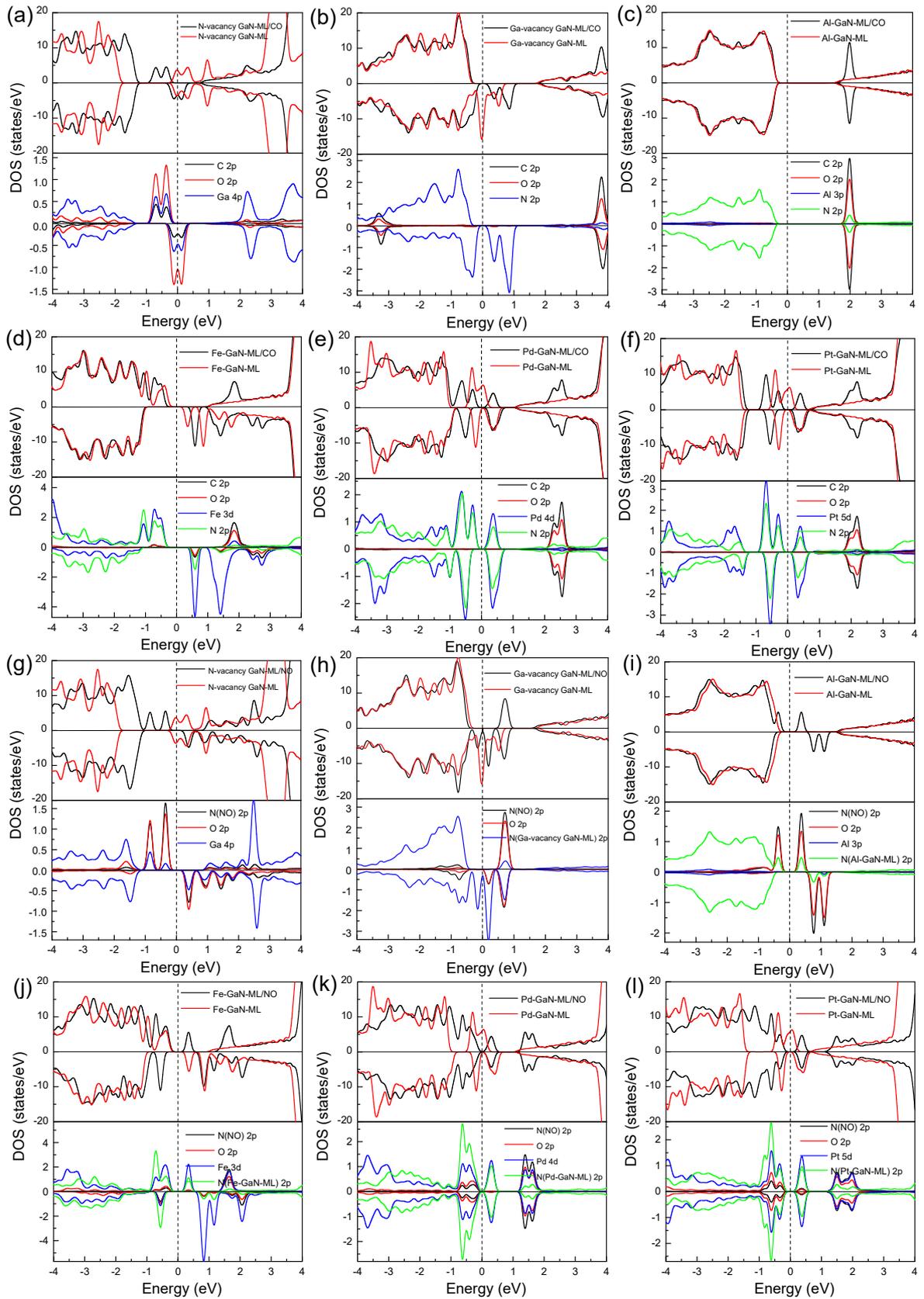

**Fig. 7.** The total DOS and PDOS of CO ((a)-(f)) and NO ((g)-(l)) adsorbed on the N-vacancy, Ga-vacancy, Al, Fe, Pd and Pt doped GaN-ML based on the most stable configuration. The Fermi level is set to zero energy and indicated by the black vertical dashed lines.

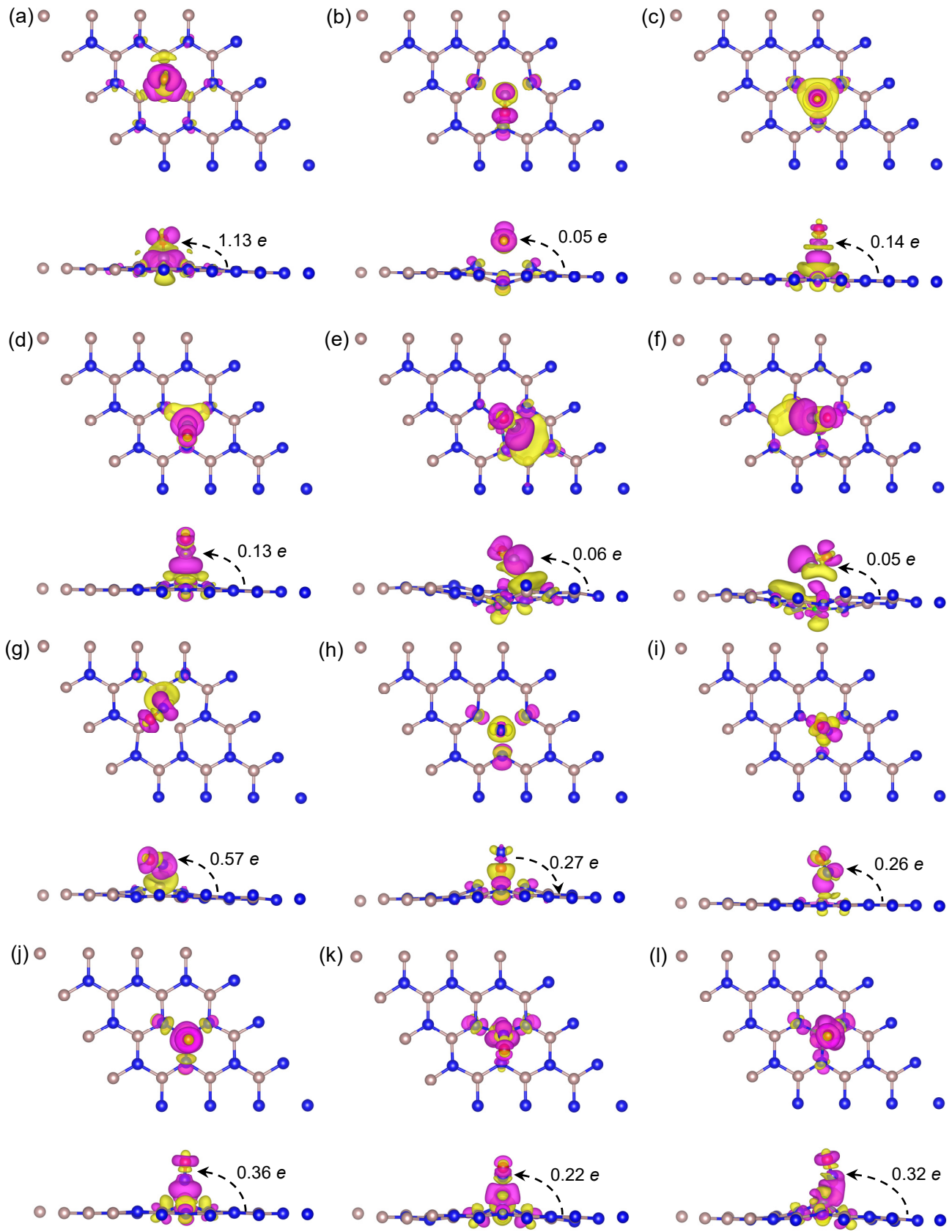

**Fig. 8.** Top and side views of the charge density difference for CO ((a)-(f)) and NO ((g)-(l)) adsorbed on the N-vacancy, Ga-vacancy, Al, Fe, Pd and Pt doped GaN-ML based on the most stable configuration. The prunosus and yellow regions correspond to charge accumulation and depletion, respectively. The isosurface is taken as 0.003 e/Å$^3$ for (a) and (g)-(l); 0.001 e/Å$^3$ for (b)-(d); 0.0002 e/Å$^3$ for (e) and (f).